\documentclass[11pt]{article}

\usepackage[colorlinks,linkcolor=blue,citecolor=blue,urlcolor=blue,bookmarks,bookmarksnumbered]{hyperref}
\usepackage{amsmath,amssymb,amsfonts}
\usepackage{slashed,extarrows,bigints,cancel}
\usepackage[paper=letterpaper,margin=1.0in]{geometry}
\usepackage{graphicx,xcolor}

\usepackage{cite}
\usepackage{setspace} 

\newcommand{\be}{\begin{equation}}
\newcommand{\ee}{\end{equation}}
\newcommand{\bea}{\begin{eqnarray}}
\newcommand{\eea}{\end{eqnarray}}
\newcommand{\ba}{\begin{array}}
\newcommand{\ea}{\end{array}}
\newcommand{\ben}{\begin{enumerate}}
\newcommand{\een}{\end{enumerate}}
\newcommand{\bi}{\begin{itemize}}
\newcommand{\ei}{\end{itemize}}
\newcommand{\bc}{\begin{center}}
\newcommand{\ec}{\end{center}}
\newcommand{\bfig}{\begin{figure}}
\newcommand{\efig}{\end{figure}}
\newcommand{\bq}{\begin{quotation}}
\newcommand{\eq}{\end{quotation}}
\newcommand{\bt}{\begin{table}}
\newcommand{\et}{\end{table}}
\newcommand{\btab}{\begin{tabular}}
\newcommand{\etab}{\end{tabular}}
\newcommand{\bs}{\begin{slide}}
\newcommand{\es}{\end{slide}}

\newcommand{\IR}{\mathbb{R}}

\newcommand{\beq}{\begin{eqnarray}}
\newcommand{\eeq}{\end{eqnarray}}
\newcommand{\beqn}{\begin{eqnarray}}
\newcommand{\eeqn}{\end{eqnarray}}

\let\ba=\overline

\def\IR{\relax\leavevmode{\rm I\kern-.18em R}}
\def\ZZ{\relax\leavevmode
       \ifmmode\mathchoice
       {\hbox{\sf Z\kern-.4em Z}}
       {\hbox{\sf Z\kern-.4em Z}}
       {\lower.9pt\hbox{\scriptsize\sf Z\kern-.36em Z}}
       {\lower1.2pt\hbox{\tiny\sf Z\kern-.36em Z}}
       \else{\sf Z\kern-.4em Z}\fi}

\def\resetby#1#2{\@addtoreset{#2}{#1}}
\def\seceq{\@addtoreset{equation}{section}
              \def\theequation{\thesection.\arabic{equation}}}

\def\Label#1{\label{#1}%
                \smash{\hbox to0pt{\raise1ex\hbox{\tiny[#1]}\hss}}}
\def\noLabels{\let\Label=\label}

\def\tx{\tilde{x}}

\begin{document}

\bc

\vskip 1.0cm

\centerline{\Large \bf On the Inevitable Lightness of Vacuum}
\vskip 0.5cm
\vskip 1.0cm

\renewcommand{\thefootnote}{\fnsymbol{footnote}}

\centerline{{\bf
Laurent Freidel${}^{1}$\footnote{\tt lfreidel@perimeterinstitute.ca},
Jerzy Kowalski-Glikman${}^{2,3}$\footnote{\tt jerzy.kowalski-glikman@uwr.edu.pl (Corresponding Author)},
Robert G. Leigh${}^{4}$\footnote{\tt rgleigh@illinois.edu}
and
Djordje Minic${}^{5}$\footnote{\tt dminic@vt.edu  }
}}

\vskip 0.5cm

{\it
${}^1$ Perimeter Institute for Theoretical Physics, 31 Caroline St. N., Waterloo ON, Canada\\
${}^2$ Institute for Theoretical Physics, University of Wroclaw,\\ Pl. Maksa Borna 9, 50-204 Wroclaw, Poland\\
${}^3$National Centre for Nuclear Research, Pasteura 7, 02-093 Warsaw, Poland \\
${}^4$ Illinois Center for Advanced Studies of the Universe \& Department of Physics,\\
			University of Illinois, 1110 West Green St., Urbana IL 61801, U.S.A.\\
${}^5$Department  of Physics, Virginia Tech, Blacksburg, VA 24061, U.S.A. \\
${}$ \\
}

\ec

\vskip 1.0cm

\begin{abstract}
In this essay, we present a new understanding of the cosmological constant problem, 
built upon the realization that the vacuum energy density can be expressed in terms of a phase space volume.  We introduce a UV-IR regularization which implies a relationship between the vacuum energy and entropy. 
Combining this insight with the holographic bound on entropy then yields a bound on the cosmological constant consistent with observations.
It follows that the universe is large, and
the cosmological constant is naturally small, because the universe is filled with a large number of degrees of freedom. 
\end{abstract}

\vspace{1cm}

\begin{center}
Essay written for the Gravity Research Foundation\\ 2023 Awards for Essays on Gravitation.\\
Submission date: March 30, 2023.
\end{center}

\renewcommand{\thefootnote}{\arabic{footnote}}

\newpage

The cosmological constant problem is considered to be one of the most important open problems in theoretical physics,
especially given the successes of Einstein's general theory of relativity
and the Standard Model of particle physics and of cosmology \cite{Weinberg:1988cp, Weinberg:2000yb, Witten:2000zk, Padmanabhan:2002ji, Polchinski:2006gy}.
The sense of urgency regarding this problem has increased 
since the discovery of dark energy \cite{Riess:1998cb, Perlmutter:1998np}
which can be translated into a small and positive cosmological constant.
The measured value of the cosmological constant differs from the canonical estimates of vacuum energy
in the context of quantum field theory \cite{Weinberg:1988cp, Weinberg:2000yb, Witten:2000zk, Padmanabhan:2002ji, Polchinski:2006gy} by many orders of magnitude.
The cosmological constant problem, or the vacuum energy problem, is also associated with the enormous hierarchy of scales
between the observed vacuum energy scale and the naive quantum gravity scale set by the Planck energy. 

In the absence of gravity, effective field theory (EFT) methods are expected to apply and in fact are foundational in 
the way that we think of quantum field theoretic physics, including
the EFT of gravity and cosmology. It is these methods that lead to the wrong result for the cosmological constant.
Moreover, it is without a doubt that concepts such as gravitational entropy, holography and related quantum information theoretic ideas are important ingredients in our understanding of gravity and are expected to play a central role in a quantum theory of gravity \cite{Harlow:2022qsq}. Therefore, it is natural to ask if these concepts are important for the central gravitational hierarchy problem, that of the vacuum energy density, and if they complicate the application of EFT methods \cite{Cohen:1998zx}, \cite{Donoghue:2020hoh}.

There are many influential reviews on the vacuum energy density problem 
\cite{Weinberg:1988cp, Weinberg:2000yb, Witten:2000zk, Padmanabhan:2002ji, Polchinski:2006gy}
in which various approaches to the cosmological constant problem have been summarized.
In this essay we present a new understanding of the cosmological constant problem based on our recent work \cite{Freidel:2022ryr}.
Our approach 
relies on a phase-space-like understanding of the vacuum energy, 
combined with the essential properties of gravitational entropy. According to the calculation presented in what follows, 
the vacuum energy and the cosmological constant are naturally small, because the universe is filled with a large number of degrees of freedom. 
In a nutshell we show that the cosmological constant size is inversely proportional to the number of degrees of freedom. Hence a universe with very few degrees of freedom would behave as effective field theory predicts, but a full one necessarily has to possess a small cosmological constant. 
This  connection between the size of the universe and the counting of states in it requires a fundamental Ultraviolet-Infrared (UV-IR) connection which we now describe in more detail. 

We start our presentation with a summary of the textbook evaluation \cite{Polchinski:1998rq, Polchinski:1985zf}
of the cosmological constant from the canonical expression for the vacuum energy which 
represents a sum (integral) over momentum space labels weighted by the canonical quantum expression for the zero point energy
\be\label{textbookrho}
\rho = \sum_k \frac{1}{2} \omega_k \to \rho = \int \frac{d^3 k }{(2\pi)^3}\frac{1}{2} \omega_k,
\ee
where $\omega_k = \sqrt{\vec{k}^2 +m^2}$ for a scalar field $\phi(x)$.\footnote{ Here we work in units where $c=\hbar=1$ therefore $\rho= E/V_3=1/V_4=\Lambda_4$. In reference \cite{Freidel:2022ryr} the reader can find the factors of $\hbar$ restored.} (The following discussion
could be generalized for other fields as well.)
This expression for the vacuum energy is naively quartically divergent in four spacetime dimensions and thus it has to be regulated.
That is, 
the regularized vacuum energy density scales as the volume
of energy-momentum space $\Lambda_4$ (in what follows we consider the case of four spacetime dimensions, even though our discussion applies
more generally)
\be
\rho \sim \Lambda_4 .
\ee
When multiplied by the Newton gravitational
constant (as implied by Einstein's equations of general relativity that include the cosmological constant term), this regulated expression, usually cut off by the Planck scale, would give a huge cosmological constant \cite{Weinberg:1988cp, Weinberg:2000yb, Witten:2000zk, Padmanabhan:2002ji, Polchinski:2006gy}
\be
\Lambda^{\mathrm{EFT}}_{cc} \sim 8\pi G_N \rho \sim \rho \ell_P^2\sim m_P^2,
\ee
where $G_N \sim \ell_P^2$ is the four dimensional Newton's gravitational constant and $\ell_P$ is the corresponding Planck length.
This problem arises both in quantum field theory and in a quantum theory of gravity and matter, such as string theory  \cite{Polchinski:1998rq, Polchinski:1985zf}.
However, the observed cosmological constant is positive and small \cite{Riess:1998cb, Perlmutter:1998np}.
(Arguments like supersymmetry do not help with the offending scaling of the vacuum energy with the volume of energy-momentum space,
once supersymmetry is broken \cite{Polchinski:1998rq, Polchinski:1985zf}. For the empirically unrealistic unbroken supersymmetry we get either the flat spacetime
with a zero cosmological constant, or an AdS space with a negative cosmological constant, in contradiction with observations \cite{Berglund:2022qsb}.)

One of the fundamental ways to define  the energy density is to 
use that the logarithm of the partition function of the field theory computes the free energy times the inverse temperature understood as a Euclidean time. Moreover, the free energy divided by the spatial volume is the energy density. This means that  the 
 vacuum partition function is given by 
$Z= \langle 0|\exp(-i\hat{H}T) |0\rangle= \exp( \rho V_4) $ \cite{Polchinski:1998rq}. This can also be established as a one-loop calculation where we use that the sum of vacuum fluctuation diagrams exponentiates into the connected sum involving just the one-loop integral $Z= \exp Z_{S^1}$ given by (see  \cite{Polchinski:1998rq, Polchinski:1985zf})\footnote{Note that the exponentiation  takes into account multi-particle states, so at one loop $Z_{S^1}$, and thus $\rho$,  is concerned with single particle states only.}
\be
Z_{S^1} = V_4 \int \frac{d^4 k }{(2\pi)^4} \int \frac{d\tau}{2 \tau} \exp[-(k^2 +m^2)\tau/2].
\ee
One notes that this expression is such that  performing the $\tau$ integration reduces the integration to a sum over on-shell states, reproducing eq. \eqref{textbookrho}.
One therefore obtains the following statement: the ratio of the vacuum partition function
for a particle with circular worldline,  $Z_{S^1}$, and the spacetime volume  
leads 
to the scaling of the vacuum energy with the volume of energy-momentum space \cite{Polchinski:1998rq, Polchinski:1985zf},
$\rho \sim \Lambda_4$.\footnote{In string theory, a similar structure is present, where $Z_{S^1}$ is replaced by  $Z_{T^2}$ (which, unlike the particle expression, is UV finite.)
}

Herein lies our central observation  \cite{Freidel:2022ryr}: the vacuum partition function $Z_{S^1}$ for a particle scales as the product of the volume of spacetime and the volume of momentum space, i.e., as the
covariant \emph{phase space volume}. It is natural to associate this volume with the counting of a number of states. In the classical particle picture, there is one state for each value of the position in spacetime and each value of 4-momentum. Naively, the number of states is, therefore, infinite. Both UV and IR regulators are needed to make the phase space volume finite, where it becomes the product of two large factors, $V_4$ and $\Lambda_4$. In quantum theory, there is a bound on how many states we can put in a given phase space cell provided by $\hbar$. {\it Therefore, the finite phase space volume corresponds to a finite number of quantum states.} In the unregulated theory,
this Hilbert space is $L^2(\mathbb{R})$. It is  infinite-dimensional and this infinity corresponds to the infinite number of spacetime points. Even if we put the system in a box which provides an IR regulator, the system is countably infinite-dimensional. In order to make this finite one also needs to 
introduce a UV regulator and also put momentum space in a box.
When both space and momentum space are compactified the Hilbert space is finite-dimensional \cite{BateWei}.
A convenient way to characterize the finite number of states $N$ is to transform to the modular polarization\footnote{The change of polarization is a unitary change of basis for the Hilbert space.} which corresponds to a tiling of phase space by modular cells of area $2\pi \hbar$ \cite{Freidel:2016pls}, and the number of states is given simply by the number of such cells. In other words,
the dimension of this Hilbert space corresponds to the counting of accessible phase space cells to the particle, so the modular regularization just described can be understood as a form of spacetime quantization.

This regularization goes beyond the standard regularization procedure of EFT, in which one bounds the volume of momentum space while keeping spacetime infinite, ignoring in effect the pairing between UV and IR regulators. The phase space regularization requires us to define two characteristic scales, the spacetime scale $\lambda$ and the momentum space scale $\varepsilon$ such that their product is the Planck constant  \cite{Freidel:2022ryr}
\be\label{hbarrelation}
\lambda \varepsilon = 2\pi.
\ee 
It is the basic postulate of quantum mechanics, and the content of the uncertainty relation, that these scales define an elementary unit phase space cell. Then to regularize the partition function $Z_{S^1}$   and in turn the vacuum energy, we first take two numbers $N_q$ and $N_p$ characterizing the size of spacetime and momentum space: $V_4 = (N_q\lambda)^4$, $\Lambda_4 = (N_p\varepsilon)^4$ and we demand 
that the total number of elementary cells is finite, which we write as $N=(N_qN_p)^4$. 
This idea has its roots in the notion of modular spacetime \cite{Freidel:2016pls}, being quantum spacetime, corresponding to a generic polarization of quantum theory.

Therefore, by using the basic fact that the product of the spacetime scale $\lambda$ and the energy-momentum scale $\varepsilon$ is the Planck constant,
we find that the product of the (regularized) spacetime volume and the energy-momentum space volume (i.e., the phase space volume) is the number
\be\label{eq6}
V_4 \Lambda_4 = N.
\ee
Now we utilize this regularization to compute the energy density  and we find the bound \cite{Freidel:2022ryr}
\beq\label{rhoboundraw}
\rho
\lesssim 
\frac{N}{V_4}.
\eeq
This  remarkable formula relates the energy density  to  the count of spacetime degrees of freedom accessible  to each particle. Even more remarkable is the fact that \eqref{eq6} is a precise notion of UV-IR mixing  that relates the spacetime volume with the size of momentum space.

Note that combining eqs. \eqref{eq6} and \eqref{rhoboundraw} yields $\rho\lesssim \Lambda_4$, which if $\Lambda_4\sim m_P^4$ leads to the usual vacuum energy conundrum. However, while eq. \eqref{hbarrelation} constrains the {\it size} of phase space cells, it does not constrain their {\it shape}, or in other words, the individual values of $\lambda$ and $\varepsilon$. So what physical principle can we use to fix these scales? Are we forced to take $\lambda=\ell_P$ and $\varepsilon=m_P$?

In order to settle this question we  propose to  identify the finite  number of cells $N$ with the gravitational entropy. That is, we propose to take into account the fundamental holographic nature of gravity.
Indeed, eq. \eqref{rhoboundraw} resembles the equation of state $\rho V \sim S$.  Therefore, this assumption means that the fundamental number of phase space cells is the same as the total number of states that quantum gravity can access. This  reinforces our initial interpretation of the modular quantization as a quantization of spacetime.
In the context of our universe, which is very close to a de Sitter universe, the holographic hypothesis means that this number is the same  as the area of the cosmological horizon in Planck units, that is \cite{Bekenstein:1980jp}
\beq\label{NS}
N \sim S_{grav} = \ell_P^{-2} \mathrm{Area}\sim (\ell/\ell_P)^2.
\eeq
This proposal identifies the characteristic length scale of the universe, the size of the cosmological horizon $\ell$, with  $N_q\lambda$. This latter identification exhibits the contextual nature of the argument. The scale $\lambda$ is not universal but instead is to be chosen in the context of the system in question.   Eq. \eqref{NS} fixes the momentum cutoff at $M=2\pi /\sqrt{\ell\ell_P}$. 
Given some mild assumptions ($N_p\sim 1$), this can be translated into a fixed value for $\lambda$ of order $\sqrt{\ell\ell_P}$, the geometric mean of UV ($\ell_P$) and IR ($\ell$) scales.

Now we make use of relations \eqref{rhoboundraw} and \eqref{NS} to find the bound on the cosmological constant, to wit,
\begin{align}
    \Lambda_{cc} = \rho\,
    \ell_{P}^{2} \lesssim \frac{1}{\ell^2}. 
\end{align}
Thus we find that the vacuum energy contribution to the cosmological constant is bounded from above by the size of the cosmological horizon, in spectacular agreement with observations.
If we take $\ell_P \sim 10^{-35}m$ and $\ell \sim 10^{27}m$,
which leads to the observed scale for the cosmological constant, 
we obtain a large number of phase space (modular spacetime) boxes $N \sim \ell^2/\ell_P^2 \sim  10^{124}$.

Our reasoning demonstrates that the cosmological constant is small simply because the number of phase space cells $N$ is large.  This leaves us with a simple question: why does $N$ have to be large? The first answer is of course phenomenological: the number of physical degrees of freedom in the universe is necessarily bounded by the number $N$ of phase space cells. Therefore, a physical universe like ours, which contains a large number  of degrees of freedom, is necessarily large. This is almost tautological: a nearly empty universe, corresponding to a small $N$, would have an extremely large cosmological constant and therefore be of Planckian size. 
Still, one can wonder if there is a more fundamental argument that favors a fundamentally large number of phase space cells.
We can argue that  $N$ has to be large so that the Universe is
semi-classical: the relative fluctuation  for the size of an object built out of $N$ independent units 
scales as the inverse of the square root of the number of units $\Delta\ell/\ell \sim {1}/{\sqrt{N}}$.
This is an argument  well known
from probability theory and statistical physics  which was used by Schr\"{o}dinger  \cite{erwin}
to explain
why atoms are so small compared to macroscopic bodies.

To summarize, we have argued that the Universe is large because it needs to contain lots of degrees of freedom. By the holographic principle, we expect that the total number of degrees of freedom is bounded by the cosmological area in Planck units. Moreover, as explained above, this number must be large because it is only in this case that the size  fluctuations are small. This is a way of saying that we have a large semi-classical spacetime.

\bigskip

\noindent
{\bf Acknowledgments:}
We thank  P. Berglund, P. Draper, A. Geraci, E. Guendelman, J. Heckman, T. H\"{u}bsch, T. Kephart, D. Mattingly, A. Mazumdar, 
L. McAllister, H. P\"{a}s, P. Ramond,
D. Stojkovic and T. Takeuchi for
discussions. DM and JKG thank  Perimeter  Institute  for  hospitality.
RGL is supported in part by the U.S. Department of Energy contract DE-SC0015655 and DM by the U.S. Department of Energy under contract DE-SC0020262.
For JKG, this work was supported by funds provided by the National Science Center, project number  2019/33/B/ST2/00050.
Research at Perimeter Institute for Theoretical Physics is supported in part by the Government of Canada through NSERC and by the Province of Ontario through MRI. This work contributes to the European Union COST Action CA18108 {\it Quantum gravity phenomenology in the multi-messenger approach.}

\end{document}